\input{psfig.sty}
\input{epsf.sty}

\def\mev{\,{\rm Me\kern-0.1em V}}
\def\gev{\,{\rm Ge\kern-0.1em V}}

\documentstyle[12pt]{article}

\begin{document}
\begin{center}
{\Large{\bf  Matching Current Correlators in Lattice QCD\\
to Chiral Perturbation Theory}}\\
\vspace*{.45in}
{\large{A. ~Duncan$^1$,
S. ~Pernice$^2$ and
J.~Yoo$^1$}} \\ 
\vspace*{.15in}
$^1$Dept. of Physics and Astronomy, Univ. of Pittsburgh, 
Pittsburgh, PA 15260\\
$^2$Engineering Department, Universidad del CEMA, Buenos Aires, Argentina
\end{center}
%%%%%%%%%%%%%%%%%%%%%%%%%%%%%%%
\vspace*{.3in}
\begin{abstract}
Chiral perturbation theory gives direct and unambiguous
predictions for the form of various two-point hadronic
correlators at low momentum in terms of a finite set of couplings
in a chiral Lagrangian. In this paper we study the feasibility
of extracting the couplings in the chiral Lagrangian (through
1-loop order) by fitting two-point correlators computed in
lattice QCD to the predicted chiral form. The correlators
are computed using a pseudofermion technique yielding all-point
quark propagators which allows the computation of the full
four-momentum transform of the two-point functions to be obtained
without sacrificing any of the physical content of the unquenched gauge
configurations used. Results are given for an ensemble of dynamical configurations
generated using the truncated determinant algorithm on a large coarse
lattice. We also present a new analysis of finite volume effects
based on a finite volume dimensional regularization scheme which preserves
the power-counting rules appropriate for a chiral Lagrangian.
\end{abstract}

\section{Introduction}

  The introduction of the concept of an effective chiral Lagrangian \cite{wein}, and
the subsequent development of systematic chiral perturbation theory \cite{Leut}, have
led to remarkable clarifications in our understanding of the properties of 
quantum chromodynamics (QCD) in the low-momentum regime. Much of the phenomenology
of the interactions of hadrons at low energies can be understood and organized using chiral
theory, where the basic degrees of freedom are colorless light hadrons, rather than
the quarks and gluons appropriate for a description at (presumably arbitrarily)
short distance scales. Of course, the parameters in a chiral effective Lagrangian
are ultimately determined by the underlying microscopic theory, i.e. the QCD
Lagrangian. The connection between the underlying theory and the chiral Lagrangian 
is however completely nonperturbative, so any determination of chiral parameters
from QCD must have recourse to a systematic nonperturbative calculational procedure:
at the present time, this means lattice gauge theory. 

  Perhaps the most direct and unambiguous predictions of chiral perturbation theory
refer to the low-momentum behavior of various two-point correlators of hadronic 
densities and currents. Typically, the low-momentum expansion of these correlators
is ordered in a series of increasing powers of the momentum-squared $p^2$, with 
the power-counting rule $m_{\pi}^2 \simeq p^2$. The leading order term is then 
determined by the lowest-dimension part ${\cal L}_2$ of the full chiral Lagrangian
${\cal L}_{\rm chir}={\cal L}_2+{\cal L}_4+...$ treated at tree level, the next to
leading term at low-momentum in hadronic correlators is obtained from use of ${\cal L}_2$
at {\em one-loop} level, together with ${\cal L}_4$ at tree-level, and so on. In 
this paper we show that the accurate  extraction of up to the leading three terms of this
expansion is perfectly feasible by simulation of the relevant correlators in lattice
gauge theory. We use exclusively unquenched configurations throughout, so there are no
issues of quenched chiral logs, for example. However, the great expense needed to
generate decorrelated dynamical configurations at light quark mass (our configurations 
were generated using the truncated determinant algorithm (TDA) \cite{TDA}) 
makes it obviously desirable to extract the maximum possible physical content from
each gauge configuration. Consequently, we have used a stochastic  pseudofermion
technique \cite{Lat01},\cite{Michael}  to obtain all-point quark propagators for each gauge configuration.
This means that the Fourier transform used to determine each two-point correlator studied
\begin{equation}
    \Delta(p) = \sum_{x,y} e^{ip\cdot(x-y)}<O(x)O(y)>
\end{equation}
contains a double sum over all points on the lattice and therefore a factor of the
lattice volume more terms than the equivalent correlator studied with single-source
or single-sink quark propagators. 

 Another aspect of  calculations performed on discrete space-time lattices is the
necessary presence of an infrared (large-distance) cutoff. In this paper we present 
results for hadronic correlators on a physically large
but coarse lattice (6$^4$ lattice with lattice spacing $a$=0.4 F, O(a$^2$) improved
gauge action) at two different sea-quark masses (note: in all the calculations 
the valence and sea-quark masses are kept the same).
Using a finite volume generalization of dimensional
regularization, we have computed the appropriate finite volume modifications of the
chiral formulas of Ref.\cite{Leut}. Of course, the systematic errors induced by 
lattice discretization will require availabity of a variety of unquenched lattices at
differing couplings.

  The presentation of our results in this paper is organized as follows. In Section 2,
 we review the all-point pseudofermion technique \cite{Lat01},\cite{Michael} used to obtain quark propagators for
 the measurement of hadronic correlators (for a more detailed study of the statistical features
 and computational cost of this method see \cite{DuncEich}). In Section 3,
 we present results for the pseudoscalar-pseudoscalar two-point function for our unquenched
 lattice ensemble. In particular, we show that consistent fits to the predicted chiral
 form can be obtained, determining the corresponding one-loop chiral Lagrangian parameters
 to a (statistical) accuracy of a few percent. Moreover, the fitted higher order coefficients are quite
 small, suggesting that chiral perturbation theory can continue to be accurate for surprisingly
 high momentum (the chiral fits are extended up to $q^2\simeq$2.5 GeV$^2$). In fact, 
 the calculations presented here of higher order chiral coefficients
 really provide a  quantitative nonperturbative  basis for estimating the
 accuracy of low order chiral perturbation theory. 
 In Section 4, we repeat this procedure for the axial vector-axial vector
 current correlator. Section 5 contains a discussion of the finite volume dimensional
 regularization technique we have used to derive systematically the finite volume corrections to the chiral
 formulas for hadronic correlators. Of course other important systematic errors, primarily
 those due to lattice discretization, can only be addressed once a much larger selection
 of dynamical lattice ensembles at various lattice spacings and with improved quark actions
 are available.

\newpage
\section{Computing Hadronic Correlators with All-point Quark Propagators}
 
 The dynamical configurations used in the present work represent a very high investment
in computational effort. Each new configuration generated using the truncated determinant
algorithm \cite{TDA} requires evaluation of several hundred low eigenvalues of the hermitian
Wilson-Dirac (or clover) operator. On a 1.5 GHz Pentium 4 processor, this takes 3 minutes for
a 6$^4$ lattice and about 80 minutes for a 10$^3$x20 lattice. Decorrelation of physical 
quantities can take anywhere from 50 to several hundred such updates. The very high cost 
of generating properly  decorrelated unquenched gauge configurations
makes it essential  to squeeze  the maximum physical information content
from each available configuration. Conventional quark propagators obtained by (say)
a conjugate gradient algorithm only provide the quark propagation amplitude from
a single source vector (either point or smeared) to any other point on the lattice:
accordingly, the computation of a two-point correlator such as Eq(1) would require
 fixing either point $x$ or $y$ (typically, at the lattice origin), which obviously
sacrifices a great deal of physical information in the gauge configuration. The results
presented in this paper will involve correlators computed from all-point quark propagators
which give the quark propagation amplitude from any point on the lattice to any other
point.

 Here we describe briefly  an approach (originally suggested by Michael and Peisa in the
context of static quark systems, \cite{Michael}) to obtaining such propagators by simulating
bosonic pseudofermion fields.
For each quark propagator needed in the hadronic observable of interest (in the case
of two-point current correlators, there are two quark propagators which must be
multiplied and traced appropriately) introduce a  bosonic pseudofermion field $\phi_{ma}$ with action
 ($m$ a lattice site, $a$ the spin-color index, $Q$ the Wilson or clover operator):
\begin{eqnarray}
   S(\phi)&=& \phi^{\dagger}Q^{\dagger}Q\phi \\
          &=& \phi^{\dagger}H^2 \phi,\;\;\;H \equiv \gamma_{5}Q = H^{\dagger}
\end{eqnarray}
For a given fixed background gauge field $A$, simulating the pseudofermion field with the
preceding action produces the following correlator ($<<O>>$ means the average of
 $O$ relative to the measure $e^{-S}$)  :
\begin{eqnarray}
   <<\phi_{ma}\phi^{*}_{nb}>>_{S(\phi)}&=& (H^{-2})_{ma,nb} \\
   <<\phi_{ma}(\phi^{\dagger}H)_{nb}>>_{S(\phi)} &=& (H^{-1})_{ma,nb}\\           
   &=&(Q^{-1}\gamma_{5})_{ma,nb}
\end{eqnarray}
Note that separate pseudofermion fields are needed for each quark propagator as averages of
four-point {\em bosonic} pseudofermion amplitudes  will produce contractions with the
wrong sign relative to the corresponding fermionic 4 quark amplitudes. However, it turns out
that the computational effort required is quite manageable, allowing us to obtain sufficiently
accurate all-point propagators with only a few times the computer time required for a conventional
conjugate-gradient evaluation of the corresponding single-source propagator. The reasons for this
are twofold:\\ 
(i) The pseudofermion average $<<....>>$  is efficiently implemented by a heat-bath update of pseudofermion fields.
 As the pseudofermion action is a Gaussian one, the statistical properties of this simulation
 are essentially trivial (gauge configuration updates, by contrast, involve an action with a highly
 nonlinear dependence on the field degrees of freedom), and contain no surprises.\\
(ii) For a fixed gauge field, most quantities decorrelate after a few pseudofermion sweeps. For low-momentum
 quantities, the presence of low eigenmodes of the hermitian Wilson operator $H$ can lead to
 long autocorrelations, but these can be handled by projecting out the corresponding low modes, which
 reduces the condition number and restores a more rapid decorrelation (see \cite{DuncEich}). However,
 even without such projections, we are able to obtain the momentum dependence of the correlators
sufficiently accurately to allow good fits to the chiral behavior.\\

 It is easy to see that  the computation of multipoint hadronic correlators involving $n$ quark propagators
 can be reduced to convolutions of $n$ pseudofermion fields, rapidly computed by 
 fast Fourier transform (FFT). (This is so both for local and smeared hadronic operators, although
 the quantities studied in this paper are exclusively local densities and currents.)
 For example, the full 4-momentum transform 
 $\Delta(q^2)\equiv\sum_{x,y}e^{iq\cdot(x-y)}\Delta(x,y)$ of  the 2-point
pseudoscalar correlator  is given by
\begin{eqnarray}
 \Delta(x,y)&=& <0|T\{\bar{\Psi}(x)\gamma_{5}\Psi(x)\;\bar{\Psi}(y)\gamma_{5}\Psi(y)\}|0> \nonumber \\
            &=& -<\rm{tr}((Q^{-1}\gamma_{5})_{xy}(Q^{-1}\gamma_{5})_{yx})> \nonumber \\
            &=& -<<\sum_{ab}\phi_{xa}(\phi^{\dagger}H)_{yb}\chi_{yb}(\chi^{\dagger}H)_{xa}>> \nonumber  \\
            &=& -<<(\phi^{\dagger}H\chi)_{yy}(\chi^{\dagger}H\phi)_{xx}>>
\end{eqnarray}
which becomes an easily evaluated fast Fourier transform of products of pseudofermion fields:
\begin{eqnarray}
\Delta(q^2)   = -<< \rm{FFT}(\chi^{\dagger}H\phi)(q)\rm{FFT}(\phi^{\dagger}H\chi)(-q)>>
\end{eqnarray}

 Our main purpose in this paper is to show that the chiral behavior of  QCD hadronic
correlators can be studied directly in momentum space and higher order chiral couplings 
extracted with high statistical accuracy. The unquenched lattices used here              
lattices are a large ensemble (800 configurations) of physically large, coarse lattices recently
used in a study of stringbreaking \cite{strbreak} (lattice spacing 0.4 Fermi, but with
O($a^2$) improvement of the gauge action \cite{Lepage}: thus, the gauge action includes the
 usual plaquette term as well as a twisted rectangle (``trt") operator), and at a sea-quark
kappa value of 0.2050. Quark eigenmodes up to 420 MeV
 are included exactly in the determinant. 
 In this lattice ensemble, the
 determinant contribution to the measure corresponds to two degenerate light quark flavors (corresponding
 to $M_{\pi}\simeq$200 MeV, where the lattice scale is fixed from stringtension measurements).

  We conclude this section by describing the computational effort required for
evaluation of the pseudofermion average (7) on a single gauge configuration.
For the 6$^4$  lattices, a single heat-bath update of the two pseudofermion
fields $\phi,\chi$ requires 0.366 sec. on a 1.5 GHz Pentium-4 processor. The
convolutions and FFT operations required to obtain the desired four-momentum field
$\Delta(q^2)$ in (8) require an additional 0.024 sec. and are performed after every
2 heat-bath updates of $\phi,\chi$. The final pseudofermion average for $\Delta(q^2)$
was obtained from 20000 measurements, corresponding to 2.1 Pentium-4 hrs.

\newpage
\section{Extracting Chiral Parameters from Pseudoscalar Correlators}

  Among the most basic predictions of chiral perturbation theory are expressions 
 for the low momentum behavior of correlators of hadronic densities and currents,
 first derived by Gasser and Leutwyler \cite{Leut}. In this section we study the
 extent to which a lattice computation of the two-point function of the isovector
 pseudoscalar density $\bar{\psi}\tau_{i}\gamma_5\psi$ (here, $\tau_i$ are the isospin
 generators for SU(2), $\psi$ the isodoublet quark field) can be used to constrain the
 parameters of the chiral effective Lagrangian. We work throughout with two flavors of 
 degenerate light dynamical quarks, so the relevant chiral group is SU(2)xSU(2), and
 we will adhere as much as possible to the notation of Ref \cite{Leut}, the results of
 which are in any event restricted to exactly this two-flavor case. 

  Defining $\tau_{\pm}\equiv \frac{1}{2}(\tau_1\pm i\tau_2)$, the correlator computed in 
 (7-8) corresponds to 
\begin{equation}
  \Delta_{PS-PS}(q^2) = \int d^{4}x <\bar{\psi}\tau_{-}\gamma_5\psi(x)\bar{\psi}\tau_{+}\gamma_5\psi(y)>e^{iq\cdot(x-y)}
\end{equation}
with the low-momentum chiral behavior (in infinite volume  Euclidean space) \cite{Leut}:
\begin{equation}
  \Delta_{PS-PS}(q^2) \simeq \frac{1}{2}(\frac{G_{\pi}^{2}}{q^2+M_{\pi}^{2}}+\frac{B^2}{2\pi^2}(l_4-h_1)+O(q^2))
\end{equation}
and with the pseudoscalar decay constant $G_{\pi}$ defined as 
\begin{equation}
  G_{\pi}\delta_{ij} \equiv <0|\bar{\psi}\tau_{i}\gamma_5\psi|\pi_{j}>
\end{equation}
The quark condensate $<0|\bar{\psi}_{i}\psi_j|0>= F_{\pi}^2 B\delta_{ij}$ fixes the constant $B$,
 and $l_4,h_1$ are couplings in the O($p^4$) chiral Lagrangian. Our object here is to determine
 the latter as accurately as possible using lattice data. The fits will also provide information
 on the size of even higher order terms (e.g. O($p^2$) in the pseudoscalar correlator),
 which is clearly relevant to the issue of accuracy of
 chiral perturbation theory through one loop order.

  We begin by describing the results obtained from our ensemble of 800 6$^4$  lattices.
 These large coarse lattice configurations are O(a$^2$) improved with respect to the gauge action,
 but the quark action is unimproved Wilson, so the results obtained will necessarily contain
 large systematic effects (relative to the corresponding continuum parameters), although we 
 shall see that the statistical errors are extremely small. The configurations were generated
 using the TDA algorithm \cite{TDA} to include virtual quark-loop  effects of a doublet of degenerate light sea quarks
 {\em exactly} up to a quark off-shellness of about  420 MeV.  The average value obtained
 for $\Delta_{PS-PS}(q^2)$ for this ensemble is shown in Fig.1. For nonzero momentum, the
 statistical errors are smaller than the symbol size.  The range of $q^2$ corresponds
 to lattice values from 0 to 16, with a unit of $q^2$ on the lattice corresponding to 0.25 GeV$^2$
 in physical units. The comparatively large error at the zero momentum point is attributable
 to the autocorrelations induced by low eigenmodes of $H$ (see \cite{DuncEich} for a  solution of
 this problem). In fact, our results for the nonzero momentum modes are sufficiently accurate
 that we will discard the zero-momentum point entirely in performing the fits to Eq(10).  

\begin{figure}
\psfig{figure=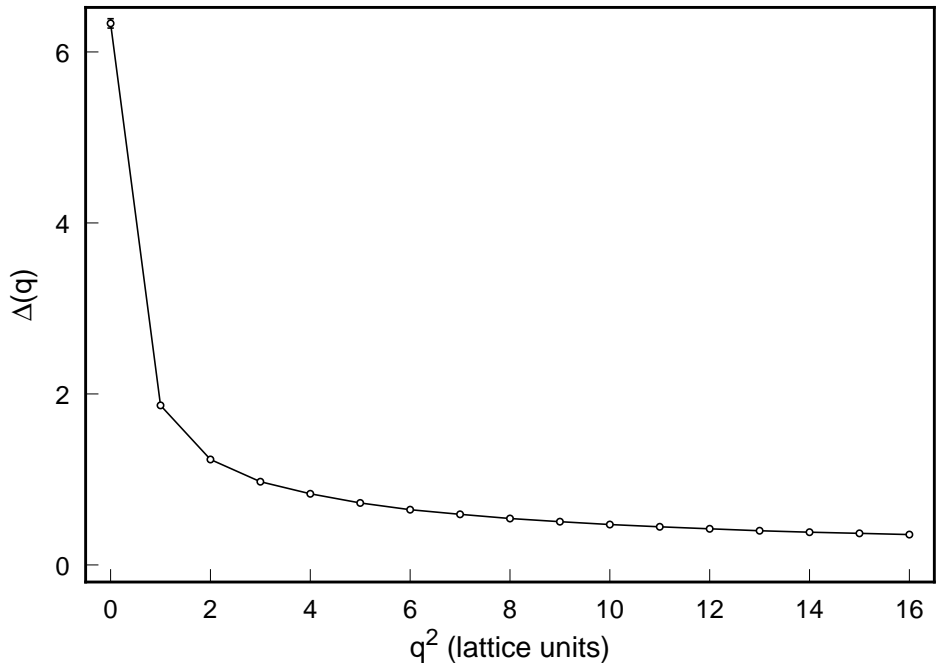,height=0.6\hsize}
\caption{Measured pseudoscalar correlator (ensemble of 800 6$^4$ lattices) $\Delta(q^2)$}
\end{figure}

\input pimass.tbl

  As the pseudoscalar correlator (9) involves by definition only local operators, it may be
 wondered whether there is much hope of extracting an accurate pion mass $M_{\pi}$ in (10).
 The excited state contamination which would normally require the use of smeared sources/sinks
 to obtain an accurate pion mass appears as contributions to the higher order chiral
 contributions (of order O($q^0,q^2,q^4$...)). A conventional cosh fit to 800 smeared-local
 correlators at $\kappa=$0.2050  gives an alternative determination of the pion mass for this ensemble:
 $M_{\pi}=0.396\pm 0.007$ (or about 200 MeV in physical units, with the scale determined from
 string tension measurements \cite{strbreak}: the use of an unimproved Wilson action on this
coarse a lattice  makes spin-dependent scales such as the rho unreliable). 
The inevitable  breakdown of chiral perturbation theory
 in the ultraviolet means that we cannot expect the fit to give meaningful results if the momentum range is
 too large. Indeed, we find that, allowing the pion mass to vary freely in a fit of the form
\begin{equation}  
  \Delta(q^2) = \frac{A_1}{q^2+M_{\pi}^2}+A_2+A_3 q^{2}+A_4 (q^{2})^{2}
\end{equation} 
the fit values for $M_{\pi}$ increase well beyond the physical value if the fit range 
is extended beyond $q^2\simeq$2.5 GeV$^2$ (see Table 1). There is a broad minimum in the
chisquared per degree of freedom for the momentum range $1\leq q^2 \leq 10-12$: choosing
 the UV cutoff in our fits at $q^2=10$ (lattice units,$\simeq2.5$Gev$^2$) gives a fitted pion mass
 very close to the value found using smeared source correlators. 

\begin{figure}
\psfig{figure=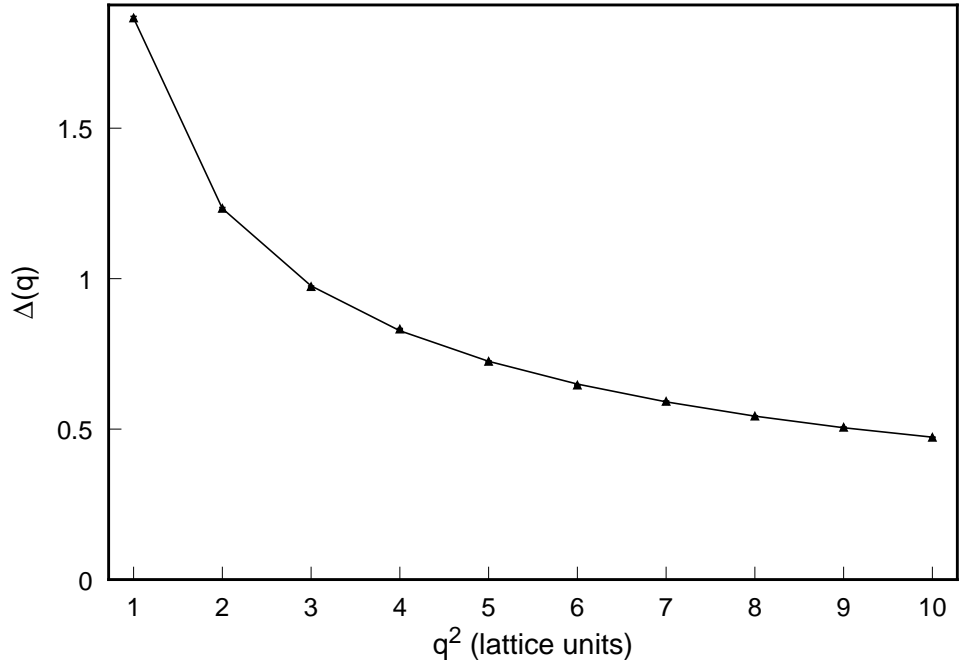,height=0.6\hsize}
\caption{Fit of measured pseudoscalar correlator $\Delta(q^2)$, $1\leq q^2\leq 10$}
\end{figure}

 A five parameter fit of the formula (12) performed on the range $1<q^2<10$ yields (see Fig. 2)
  the following results:  
\begin{eqnarray}
    A_{1}&=& 1.52 \pm 0.029 \nonumber  \\
    M_{\pi}&=& 0.422\pm 0.020 \nonumber \\
    A_{2}&=& 0.622 \pm 0.010 \nonumber \\
    A_{3}&=& -0.0460 \pm 0.0015 \nonumber \\
    A_{4}&=& 0.00163\pm 0.00007 
\end{eqnarray}
with a $\chi^2/{\rm d.o.f}=2.6$. The pion mass is consistent within errors with the value quoted above, determined
 from a study of smeared-local correlators. The remarkable statistical accuracy of this method is apparent in
 the errors for the leading order residue $A_1$ and the one-loop chiral parameter $A_2$, both of which are determined
 at the 2\% level for the statistical error! Moreover, the higher order coefficients ($O(q^2,q^4)$ terms) are {\em very} small, which bodes
 well for the accuracy of chiral perturbation theory through one-loop, even at quite high momenta.
   Repeating the fit with the pion mass held fixed at the value $M_{\pi}=$0.396 determined from fits of smeared-local
 correlators gives a somewhat improved $\chi^2/{\rm d.o.f}$=2.2, with the fitted parameters statistically 
 indistinguishable from the results in (13). However, removing the $q^4$ term in (12) results in fits
 with much higher $\chi^2/{\rm d.o.f.}>$5, and fit parameters $A_1,A_2$ and $M_{\pi}$ differing   
 substantially from the values obtained in (13) if the momentum range is extended beyond $q^2=$6, corresponding
 to 1.5 GeV$^2$ in physical units. 
    The value of the pion decay constant will be obtained from a study of the axial-vector correlator in the
 next section: we find $F_{\pi}=$0.187$\pm$0.011.  The critical kappa value is
 most readily extracted from the predicted topological charge distributions of Leutwyler and Smilga
 \cite{LeutSmil},\cite{Eich}: one finds $\kappa_c$=0.2067, corresponding
 to a bare quark mass for $\kappa=0.2050$ of $\hat{m}=$0.020. On the other hand, using the value for $F_{\pi}$
 quoted above,  the bare quark mass can be
 extracted from the chiral Ward identity
\begin{equation}
  \hat{m} = \frac{F_{\pi}M_{\pi}^2}{G_{\pi}}
\end{equation}
 where from (10,13) we find $G_{\pi}=$1.74$\pm$0.02. Accordingly (taking $M_{\pi}=$0.422), (14) yields a bare quark mass
\begin{equation}
  \hat{m} = 0.0191\pm 0.0021
\end{equation}
 consistent with the $\kappa_c$ value extracted from the topological analysis. 
     
\newpage
\section{Extracting Chiral Parameters from Axial Current Correlators}

In this section we shall describe our results for the nonperturbative lattice evaluation of the
axial vector correlator, using the same lattice ensemble described in the preceding
section. Defining
\begin{equation}
  \Delta_{AX-AX}^{\mu\nu}(q) = \int d^{4}x <\bar{\psi}\tau_{-}\gamma_5\gamma^{\mu}\psi(x)\bar{\psi}\tau_{+}\gamma_5\gamma^{\nu}\psi(y)>e^{iq\cdot(x-y)}
\end{equation}
we have used the pseudofermion technique to extract the contracted scalar quantity (in Euclidean space)
\begin{equation}
 \Delta_{AX-AX}(q^2) \equiv \frac{1}{2}g_{\mu\nu}\Delta_{AX-AX}^{\mu\nu}(q)
\end{equation}
  The prediction of chiral perturbation theory for this contracted  axial-vector correlator can be summarized in the
 formula \cite{Leut} :
\begin{equation}
  \Delta_{AX-AX}(q^2) \simeq \frac{F_{\pi}^{2}q^2}{q^2+M_{\pi}^2}-4F_{\pi}^2+\frac{1}{16\pi^2}(l_5-h_2)q^2+O(q^4)
\end{equation}
where the term involving $l_5-h_2$ represents a contribution from the O($p^4$) chiral Lagrangian. The constant term
$-4F_{\pi}^2$ in (18) arises from a contact term $\propto \delta^{4}(x-y)$ in the coordinate space correlator which is
highly UV-divergent and not accessible from our lattice calculation, so we have allowed this constant to float 
freely in our fits. 

\begin{figure}
\psfig{figure=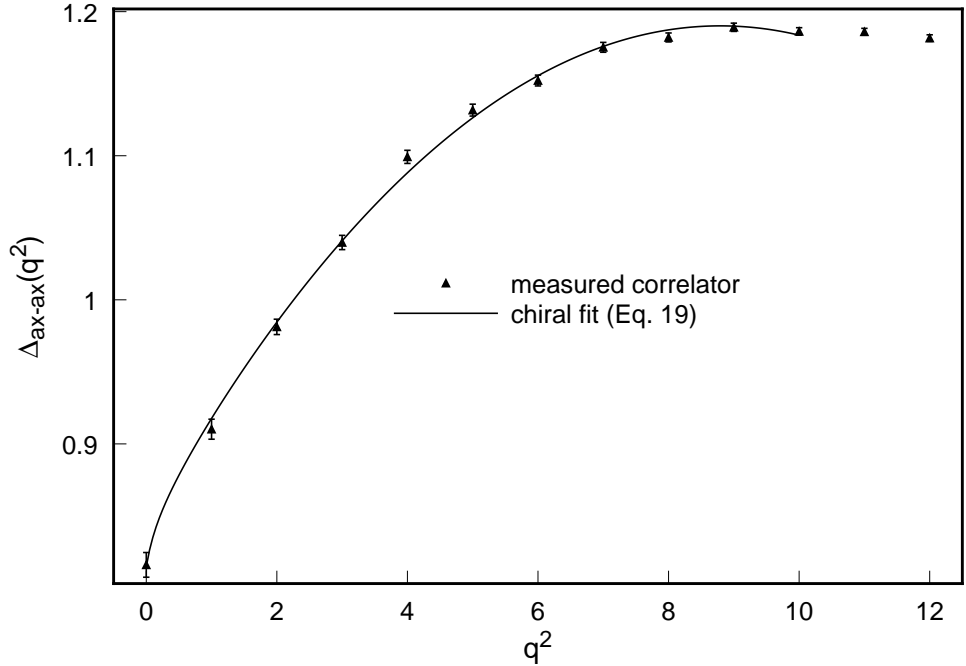,height=0.6\hsize}
\caption{Measured axial-vector correlator  $\Delta_{ax-ax}(q^2)$, fit $0\leq q^2\leq 10$}
\end{figure}

  We have computed $\Delta_{AX-AX}(q^2)$ using the pseudofermion approach: for the ensemble of 800 lattice
configurations, the results are shown in Fig. 3. Again, the statistical errors are quite small. 
With the fitting formula
\begin{equation}
  \Delta_{AX-AX}(q^2) = \frac{F_{\pi}^{2}q^2}{q^2+M_{\pi}^2}+A_{1}+A_{2}q^{2}+A_{3}q^{4}
\end{equation}
a conservative fit using only the $q^2$ range from 0 to 4 (lattice units) and setting the higher order
$A_{3}$ term to zero, one finds $F_{\pi}=$0.213$\pm$0.010 with a $\chi^{2}/{\rm dof}$ of 0.8/2. In this fit
we have fixed the pion mass at the value 0.422 determined in the preceding section from the pseudoscalar
correlator fits. The higher order coefficients are determined as $A_{1}=$0.8157$\pm$0.0040, $A_{2}=$0.0602$\pm$0.0004 respectively.
The statistical error on these higher order chiral coefficients is again very small, indicating the high
statistical content available in an all-point approach. The contact term $A_{1}$ is stable to three significant
figures with respect to changing the fitting range to $0\leq q^2\leq$3 or $0\leq q^2\leq$5, as it is completely
dominated by ultraviolet contributions (see discussion above).
Another point worth noting here is that our momentum-space approach allows the extraction of 
meaningful chiral parameters at light quark mass from a local-local axial-vector correlator, something which would
be impossible using conventional Euclidean coordinate space fits.

  Keeping only terms up to order $q^2$ in the chiral fitting formula, one finds in the case of the axial-axial correlator that 
the chisquared deteriorates much more rapidly as higher momenta are included in the fitting range than in the case
of the pseudoscalar correlators discussed in the preceding section. If a $q^4$ term is included in the fit, then the
fitting range can be extended considerably. In Fig. 3 the chiral fit obtained for a fitting range $0\leq q^2\leq$10
is shown: the $\chi^{2}/{\rm dof}$ for this fit is 14/7. The extracted chiral parameters in this case are:
\begin{eqnarray}
   F_{\pi} &=& 0.187 \pm 0.011  \nonumber \\
   A_{1} &=& 0.8148 \pm 0.0038 \nonumber \\
   A_{2} &=& 0.0777 \pm 0.0003 \nonumber \\
   A_{3} &=& -0.00442 \pm 0.00002 
\end{eqnarray}

\section{Finite Volume Effects}
\subsection{Dimensional Regularization at Finite Volume}

In order to reliably extract the parameters of the chiral
Lagrangian out of equations (12) and (19) and the values in
equation (13) and (20), we need a generalization of equations
(10) and (18) that includes finite volume effects. At the same
time this generalization should preserve the power counting rules
appropriate for chiral perturbation theory.  Here we present such
a generalization by extending the dimensional regularization scheme
to finite volume with periodic boundary conditions.  We assume our
system lives in a ${\mathcal{L}}^4$ box with the momentum $k_\mu$
taking values $\frac{2\pi}{{\mathcal{L}}} n_\mu$ where $n_\mu$ is
an integer.

We start by converting massive propagators to exponential form
using  Schwinger parameters:
\begin{equation}
\frac{1}{k^2 + m^2} = \int_0^\infty d \lambda e^{- \lambda (k^2 +
m^2 )}
\end{equation}
and replacing the infinite volume loop integrals over $k$ (where
$k$ is a d-dimensional vector) by finite sums
\begin{equation}
\mu^{4 - d} \int \frac{d^d k}{(2 \pi)^d} f(k_\mu) \rightarrow
\frac{\mu^{4 - d}}{(2 \pi)^d}(\frac{2 \pi}{{{\mathcal{L}}}})^d
\sum_{n_\mu} f (\frac{2\pi}{{\mathcal{L}}} n_\mu)
\end{equation}

Once the k-dependence is exponentiated as in (21), we will need to
evaluate sums like
\begin{equation}
\sum_{n_\mu = - \infty }^{+ \infty} e^{- \lambda (\frac{2 \pi
n_\mu}{{\mathcal{L}}})^2} \equiv \sqrt{\pi} {\mathcal{I}} (
\frac{4 \pi^2}{{\mathcal{L}}^2} \lambda)
\end{equation}
so that (22) will involve ${\mathcal{I}}^d$, where ${\mathcal{I}}$
is the function
\begin{equation}
{\mathcal{I}} (\lambda) \equiv \frac{1}{\sqrt{\pi}} \sum_{n = -
\infty }^{+ \infty} e^{- \lambda n^2}
\end{equation}
We will also need to evaluate a generalization of ${\mathcal{I}}
(\lambda)$:
\begin{equation}
{\mathcal{J}} ( \lambda, \beta ) \equiv \frac{1}{\sqrt{\pi}}
\sum_{n = - \infty }^{+ \infty} e^{- \lambda (n^2 - \beta n)}
\end{equation}
where we shall need both ${\mathcal{I}}$ and ${\mathcal{J}}$ in the physically
interesting limit $\lambda \rightarrow 0$ corresponding to large
but finite volume.

Using the Poisson sum formula one can easily write equations (24)
and (25) in a form useful to take that limit:
\begin{equation}
{\mathcal{I}} (\lambda) = \frac{1}{\sqrt{\lambda}} \sum_{m = -
\infty }^{+ \infty} e^{- \frac{\pi^2 m^2}{\lambda}} \simeq
\frac{1}{\sqrt{\lambda}} (1 + 2 e^{- \frac{\pi^2}{\lambda}}+ ...)
\end{equation}
and
\begin{equation}
{\mathcal{J}} ( \lambda, \beta ) = \frac{1}{\sqrt{\lambda}}
e^{\frac{\beta^2 \lambda}{4}} \sum_{m = - \infty }^{+ \infty} e^{-
\frac{\pi^2 m^2}{\lambda}} \cos (\pi \beta m)
\end{equation}
These equations are all we need to develop the generalization of
dimensional regularization to finite volume that preserves the
power counting rules appropriate for chiral perturbation theory.

Consider for example the quadratically divergent integral
\begin{equation}
S_2 \equiv \mu^{4-d} \int \frac{d^d k}{(2 \pi)^d} \frac{1}{k^2 +
m^2}
\end{equation}
where we work at finite volume.  Using the procedure outlined above
and changing variables we get
\begin{equation}
S_2 \equiv \frac{1}{4} \mu^{4-d} \pi^{\frac{d}{2} -2}
{\mathcal{L}}^{2-d} \int_0^\infty d \lambda e^{- \frac{m^2
{\mathcal{L}}^2}{4 \pi^2} \lambda} {\mathcal{I}}^d (\lambda)
\end{equation}
Replacing the expression (26) for ${\mathcal{I}}$ we can easily
check that the first term in the expansion is identically equal to
the infinite volume integral while the other terms correspond to
ultraviolet finite exponentially small finite volume corrections.
The divergence as $d \rightarrow 4$ is therefore equal to the
infinite volume divergence and the $MS$ or $\overline{MS}$
prescriptions remain unmodified.

If $m\mathcal{L}$ in the exponent above is large,
the next to leading term in the expansion of ${\mathcal{I}}^4$
approximates well the finite volume correction of $S_2$.  We
can make $d = 4$ since the term is finite, getting
\begin{equation}
S_2^{\mathrm{next \, to \, leading}} = \frac{2}{ {\mathcal{L}}^{2}
} \int_0^\infty d \lambda \frac{e^{- \frac{m^2 {\mathcal{L}}^2}{4
\pi^2} \lambda - \frac{\pi^2}{\lambda}}}{\lambda^2} =
\frac{2}{\pi^2} \frac{m}{ {\mathcal{L}}} K_1 (m {\mathcal{L}})
\end{equation}
where $K_1 (x)$ is a modified Bessel function.  This is
exponentially small when $m {\mathcal{L}}$ gets large. Further 
corrections fall exponentially at an even faster rate. 
For the lattice ensemble studied previously, $m{\mathcal{L}}{\sim} 2.5$
so this approximation is inadequate and (29) must be evaluated
more carefully (see below).

As another example of the finite volume dimensional regularization
procedure consider the relation
\begin{equation}
S_4 \equiv \mu^{4-d} \int \frac{d^d k}{(2 \pi)^d} = 0
\end{equation}
for the quartically divergent integral.  For infinite volume this
is the only consistent definition, as there is no available
dimensionful quantity other than $\mu$, which can only appear
logarithmically in the finite volume part at $d = 4$.  At finite
volume this argument does not work because we have now a new scale
${\mathcal{L}}$.  However, as we show next, the relation is also
valid at finite volume.

Writing $1$ in the ``integrand" as $(m^2 + k^2)/(m^2 + k^2)$ and
introducing again the Schwinger parameter we can easily transform
$S_4$ into
\begin{equation}
S_4 = \frac{\mu^{4-d} \pi^{d/2}}{{\mathcal{L}}^d} \int_0^{\infty}
d \lambda e^{-x \lambda} (x - \frac{\partial}{\partial
\lambda}){\mathcal{I}}^d (\lambda)
\end{equation}
where $x = \frac{m^2 {\mathcal{L}}^2}{4 \pi^2}$ as usual.
Separating ${\mathcal{I}}^d(\lambda)$ into $\lambda^{d/2}$ and
${\mathcal{I}}^d(\lambda)- \lambda^{d/2}$ and replacing each part
into the above equation, we get for the first part
\begin{equation}
S_4^{\mathrm{divergent}} = \frac{\mu^{4-d}
\pi^{d/2}}{{\mathcal{L}}^d} \int_0^{\infty} d \lambda e^{-x
\lambda} (x - \frac{\partial}{\partial \lambda}) \lambda^{d/2}
\end{equation}
which is identical to the infinite volume version of $S_4$ and
therefore it must be zero by the standard argument.  In fact,
analytically continuing equation (33) to $d < 0$ and integrating
by parts one can see that it is zero as it must be.

The non trivial result is that the finite volume correction
\begin{equation}
S_4^{\mathrm{convergent}} = \frac{\mu^{4-d}
\pi^{d/2}}{{\mathcal{L}}^d} \int_0^{\infty} d \lambda e^{-x
\lambda} (x - \frac{\partial}{\partial \lambda})
({\mathcal{I}}^d(\lambda)- \lambda^{d/2})
\end{equation}
is also explicitly zero for {\it any} $d$, as one can see by
replacing the expansion (26) for ${\mathcal{I}}(\lambda)$ above
and integrating by parts.

It is now clear that with the procedure outlined in equations (21)
to (27) we can compute the finite volume effects in the
dimensional regularization scheme.  In the next subsection we
shall employ this technology to
compute the finite volume corrections to the two-point hadronic
correlators. This will allow us to estimate quantitatively the
systematic errors in the results of Sections 3,4 due to finite size.

\subsection{Finite Volume Corrections to Two-Point Hadronic Correlators}

Let us find finite volume corrections to the pseudoscalar correlator
and the axial vector correlators using the dimensional techniques
described above. First, we note that the finite-volume dependence
of correlators calculated through O($p^4$) in chiral perturbation 
theory arises from the one-loop integrals using the lowest order
chiral Lagrangian for the chiral vertices (see Ref \cite{Leutfinvol}).
 Since one-loop graphs come from ${\cal L}_2$, 
the evaluation of pseudoscalar and axial vector correlators involves
loop integrals of only the type (28) \cite{Leut}. So in calculating 
the finite volume corrections to $G_\pi$, $M_\pi$, and $F_\pi$ to one
loop order we can simply add the finite correction to this loop integral
to the mass logarithm terms in those constants obtained in Ref \cite{Leut}. 

The $G_\pi$, $M_\pi$, and $F_\pi$ modifications are then given by
\begin{equation}
G_\pi \rightarrow G_\pi^{{\mathrm{inf.}} \, {\mathrm{vol.}}} -
\frac{B}{F} S_2^F
\end{equation}
\begin{equation}
M_\pi \rightarrow (M_\pi^{{\mathrm{inf.}} \, {\mathrm{vol.}}})
+ \frac{M}{4F^2} S_2^F
\end{equation}
and
\begin{equation}
F_\pi \rightarrow F_\pi^{{\mathrm{inf.}} \, {\mathrm{vol.}}} -
\frac{1}{F} S_2^F
\end{equation}
where
\begin{equation}
S_2^F \equiv \frac{1}{4} \frac{1}{{\mathcal{L}}^{2}} \int_0^\infty
d \lambda e^{- \frac{M^2 {\mathcal{L}}^2}{4 \pi^2} \lambda}(
{\mathcal{I}}^4 (\lambda) - \frac{1}{\lambda^2})
\end{equation}
which corresponds to the finite volume part of $S_2$ of equation
(29).  Since the coefficient in the exponent in the integrand,
$M^2{\mathcal{L}}^2/4\pi^2$, is small($\simeq$ 0.16), the contribution from large
$\lambda$ is substantial. 
So the original form (24) for $\mathcal{I}(\lambda)$ is used. 
The numerical integration using Mathematica
yields $S_2^F$=0.002200. To find the finite volume corrections to 
the leading order, the values of the constants $F_\pi$=0.187$\pm$0.011, 
and $M_\pi$=0.422$\pm$0.020
found in the previous sections are used
for $F$, and $M$. With $S_2^F$ above, the finite
volume correction to $M_\pi$ is 0.007 and $F_\pi$ -0.012. 
In the finite volume term of $G_\pi$there is a constant B(practically, the
quark condensate) which was not
extracted directly from the previous fits. To the leading order
$G_\pi=2BF$ \cite{Leut}. Thus the finite volume correction to $G_\pi$
becomes $-G_\pi{S_2^F}/2F_{\pi}^2$ to this order. 
Using a central value of $G_\pi$=1.74, we find -0.05 for the finite volume
correction to $G_\pi$. We see that in all cases the finite volume corrections
are on the order of a few percent even for the fairly light quark mass
($M_{\pi}\simeq$ 200 MeV) used in the dynamical lattices studied in
Sections 3,4. Of course, these lattices were physically large (2.4 F$^4$):
on smaller lattices for light quark masses, the finite volume
corrections will be more important.

\newpage

\section{Acknowledgements}
The work of A. Duncan and J. Yoo was supported in part by 
NSF grant PHY00-88946.

\end{document}